\documentclass{article}
\usepackage{amsmath}
\usepackage{latexsym}
\usepackage{graphicx}
\usepackage{color}
\usepackage{amssymb}
\usepackage{subfigure}
\usepackage{tabularx}
\usepackage[labelfont={bf,sf},tableposition=top]{caption}

\newtheorem{theorem}{Theorem}
\newtheorem{acknowledgement}{Acknowledgement}
\newtheorem{definition}{Definition}

\newtheorem{proposition}{Proposition}
\newtheorem{remark}{Remark}

\begin{document}
\title{First characterization of a new method for numerically solving the Dirichlet problem of the two-dimensional Electrical Impedance Equation}
\author{M. P. Ramirez T.$^{(1)}$, C. M. A. Robles G.$^{(1)(2)}$\\
and R. A. Hernandez-Becerril$^{(1)}$\\
\small{$^{(1)}$Communications and Digital Signal Processing Group,}\\ 
\small{Faculty of Engineering, La Salle University,}\\
\small{B. Franklin 47, C.P. 06140, Mexico.} \\ 
\small{$^{(2)}$SEPI, ESIME Culhuacan, National Polytechnic Institute,}\\
\small{Av. Santa Ana No. 1000, C.P. 04430, Mexico.}}

\date{}
\maketitle
\begin{abstract}
Based upon elements of the modern Pseudoanalytic Function Theory, we analyse a new method for numerically approaching the solution of the Dirichlet boundary value problem, corresponding to the two-dimensional Electrical Impedance Equation. The analysis is performed by interpolating piecewise separable-variables conductivity functions, that are eventually used in the numerical calculations in order to obtain finite sets of orthonormal functions, whose linear combinations succeed to approach the imposed boundary conditions. To warrant the effectiveness of the numerical method, we study six different examples of conductivity. The boundary condition for every case is selected considering one exact solution of the Electrical Impedance Equation. The work intends to discuss the contributions of these results into the field of the Electrical Impedance Tomography.
\end{abstract}
\section{Introduction}

The study of the Dirichlet boundary value problem for the two-dimensional Electrical Impedance Equation
\begin{equation}
\mbox{div}(\sigma\mbox{grad}u)=0,
\label{int00}
\end{equation}
is fundamental for well understanding its inverse problem, usually know as Electrical Impedance Tomography, and that was first posed in correct mathematical form by A. P. Calderon \cite{calderon} in 1980. It is remarkable that for more than twenty years after the problem was stated, the mathematical complexity of (\ref{int00}) could provoke that many experts considered \emph{impossible} to obtain its general solution in analytic form \cite{webster}, even for the simplest cases of $\sigma$. But the this conception turned around in 2005, when V. Kravchenko \cite{kra2005} noticed that the two dimensional case of (\ref{int00}) was completely equivalent to a special kind of Vekua equation \cite{vekua}. Only one year latter, K. Astala and L. P\"aiv\"arinta \cite{astala} rediscovered independently this relation, and with it, they gave a positive answer to the Calderon's problem in the plain.

Many other important results were obtained soon after. As an example, Kravchenko \emph{et al.} published in 2007 what can be considered the first general solution of (\ref{int00}) in analytic form \cite{kr-o}, proposed for a certain class of conductivities $\sigma$, and employing Taylor series in formal powers \cite{bers}. But we would like to claim now special attention on a recent and relevant discovering: The proof of the completeness of the set of formal powers in bounded domains, for approaching solutions in the plain of the Dirichlet problem for the Electrical Impedance Equation \cite{cck}. This will be valid when the conductivity function can be expressed by means of a separable-variables function.

On the light of all these notorious advances, we want to emphasize that it is not clear yet how to fully apply the elements of the modern Theory of Pseudoanalytic Functions \cite{kpa} into more specific, but still very important, Engineering applications, as it is the Electrical Impedance Tomography from the point of view of Medical Imaging.

The main objective of this work is to start a discussion in this direction. Even it is not clear if the proof provided in \cite{cck} can be extended for the cases when $\sigma$ is a separable-variables function, but piecewise-defined within a bounded domain, the numerical calculations show that the techniques employed in purely mathematical problems, could well be used for analysing some physical cases, providing quite acceptable results.

Basically, our work suggest that, if the values of the electrical conductivity are known at every point within a bounded domain, it will be always possible to introduce a piecewise separable-variables function that approaches the conductivity, and in consequence, to obtain a numerical orthonormal set of functions that seems to be capable of approaching solutions for the Dirichlet boundary value problem of (\ref{int00}). This would be true for a certain class of bounded domains defined on the plain, but wide enough to include most relevant cases for Experimental Physics.

We base our assessments onto a set of conductivity examples that tries to be as diverse as possible from the mathematical point of view. Possessing for each example an analytic solution, we introduce their corresponding piecewise separable-variables conductivity functions, and employing them for the numerical calculations, we approach their corresponding analytic solutions valued at the boundary. Then, a variety of qualitative and quantitative references are given, in order to estimate the effectiveness of the approach.

We make special emphasis in those examples whose conductivity functions are not separable-variables by definition, showing that, at least for the analysed cases, the technique can be trustful. This would be an important step for performing experiments with other classes of conductivities, as those emerging from geometrical distributions, whose exact mathematical expressions are, in general, unknown. If our appreciations are correct, we would be one step closer for fully applying the modern Pseudoanalytic Function Theory into Medical Imaging.

\section{Preliminaries}

Following \cite{bers}, let the pair of complex-valued functions $F$ and $G$ fulfil the condition
\begin{equation}
\mbox{Im}(\overline{F}G)>0,
\label{pre00}
\end{equation}
where $\overline{F}$ represents the complex conjugation of $F$: $\overline{F}=\mbox{Re}F-i\mbox{Im}F$, and $i$ is the standard imaginary unit $i^{2}=-1$. Therefore, any complex-valued function $W$ can be written by means of the linear combination of the pair $(F,G)$:
\[
W=\phi F+\psi G,
\]
where $\phi$ and $\psi$ are purely real-valued functions. Based upon this idea, L. Bers introduced the concept of the $(F,G)$-derivative of $W$ as
\begin{equation}
\partial_{(F,G)}W=\left( \partial_{z}\phi \right) F+\left( \partial_{z}\psi \right) G.
\label{pre01}
\end{equation}
But this derivative will exist if and only if the following condition holds:
\begin{equation}
\left( \partial_{\overline{z}}\phi \right)F+\left( \partial_{\overline{z}}\psi\right)G=0.
\label{pre02}
\end{equation}

From hereafter, we will consider $\partial_{z}=\partial_{x}-i\partial_{y}$ and $\partial_{\overline{z}}=\partial_{x}+i\partial_{y}$. Nevertheless these operators are usually introduced with the factor $\frac{1}{2}$, in this work will be somehow more convenient to work without it.

By introducing the notations
\begin{eqnarray}
A_{(F,G)}=\frac{\overline{F}\partial_{z}G-\overline{G}\partial_{z}F}{F\overline{G}-G\overline{F}},\ \ \ a_{(F,G)}=-\frac{\overline{F}\partial_{\overline{z}}G-\overline{G}\partial_{\overline{z}}F}{F\overline{G}-G\overline{F}}, \nonumber \\
B_{(F,G)}=\frac{F\partial_{z}G-G\partial_{z}F}{F\overline{G}-G\overline{F}},\ \ \ b_{(F,G)}=-\frac{G\partial_{\overline{z}}F-F\partial_{\overline{z}}G}{F\overline{G}-G\overline{F}};
\label{pre03}
\end{eqnarray}
the $(F,G)$-derivative of $W$, presented in (\ref{pre01}), can be written as
\begin{equation}
\partial_{(F,G)}W=\partial_{z}W-A_{(F,G)}W-B_{(F,G)}\overline{W},
\label{pre04}
\end{equation}
whereas the condition (\ref{pre02}) will turn into
\begin{equation}
\partial_{\overline{z}}W-a_{(F,G)}W-b_{(F,G)}\overline{W}=0.
\label{pre05}
\end{equation}

A pair of complex functions $(F,G)$ satisfying (\ref{pre00}), will be called a \emph{generating pair}, and the notations introduced in (\ref{pre03}) will be referred as the \emph{characteristic coefficients} of the generating pair $(F,G)$. Moreover, the expression (\ref{pre05}) is known as the \emph{Vekua equation} \cite{vekua}, and soon it will be seen that it possesses special significance for this work. It is also necessary to mention that every function $W$, solution of (\ref{pre05}), will be called $(F,G)$\emph{-pseudoanalytic}.

The following statements were originally posed in \cite{bers}. We present them here slightly modified in behalf of better explaining our results.

\begin{theorem}
\label{preFG}
The elements $F$ and $G$, belonging to the generating pair (\ref{pre00}), are both $(F,G)$-pseudoanalytic (\ref{pre05}), and their $(F,G)$-derivatives (\ref{pre04}) vanish identically
\[
\partial_{(F,G)}F\equiv\partial_{(F,G)}G\equiv0.
\]
\end{theorem}
\begin{theorem}
\label{theoremFG}
\cite{bers}\cite{kpa} Let $p$ be a non-vanishing function within some domain $\Omega$, and let
\begin{equation}
F=p,\ \ \ G=\frac{i}{p}.
\label{pre06}
\end{equation}
It is easy to verify that $F$ and $G$ constitute a generating pair (\ref{pre00}), whose characteristic coefficients (\ref{pre03}) are
\begin{eqnarray}
A_{(F,G)}=a_{(F,G)}=0,\nonumber \\
B_{(F,G)}=\frac{\partial_{z}p}{p},\ \ \ b_{(F,G)}=\frac{\partial_{\overline{z}}p}{p}.
\label{pre07}
\end{eqnarray}
Therefore, the corresponding Vekua equation (\ref{pre05}) of an $(F,G)$-pseudoanalytic function $W$ will have the form
\begin{equation}
\partial_{\overline{z}}W-\frac{\partial_{\overline{z}}p}{p}\overline{W}=0.
\label{pre08}
\end{equation}
\end{theorem}
\begin{definition}
Let $(F_{0},G_{0})$ and $(F_{1},G_{1})$ be two generating pairs of the form (\ref{pre06}), and let their characteristic coefficients fulfil the condition
\[
B_{(F_{0},G_{0})}=-b_{(F_{1},G_{1})}.
\]
The generating pair $(F_{1},G_{1})$ will be then called a successor pair of $(F_{0},G_{0})$, as well $(F_{0},G_{0})$ will be named a predecessor pair of $(F_{1},G_{1})$.
\end{definition}
\begin{definition}
Let the elements of the set
\begin{equation}
\left\lbrace\left(F_{m},G_{m}\right)\right\rbrace;\ \ m=0,\pm 1,\pm 2,...
\label{pre09}
\end{equation}
be all generating pairs, and let every $(F_{m+1},G_{m+1})$ be a successor $(F_{m},G_{m})$. Hence, the set (\ref{pre09}) will be called a generating sequence. Particularly, if $(F,G)=(F_{0},G_{0})$, we will say that $(F,G)$ is embedded into (\ref{pre09}). Moreover, if there exist a number $k$ such that $(F_{m+k},G_{m+k})=(F_{m},G_{m})$, we will say that the generating sequence (\ref{pre09}) is periodic, with period $k$.
\end{definition}

L. Bers also introduced the concept of the $(F,G)$-integral of a complex function $W$. The complete and detailed conditions for its existence can be found in \cite{bers} and in \cite{kpa}, but they are out of the scope of this work. Because of this, the elements we present in the upcoming paragraphs are only basic statements for supporting our results.

\begin{definition}
Let $(F_{0},G_{0})$ be a generating pair with the form (\ref{pre06}). Its adjoint pair $(F_{0}^{*},G_{0}^{*})$ will be defined as
\[
F_{0}^{*}=-iF_{0},\ \ \ G_{0}^{*}=-iG_{0}.
\]
\end{definition}
\begin{definition}
The $(F_{0},G_{0})$-integral of a complex function $W$ (if it exists) is defined according to the expression
\[
\int_{\Lambda}Wd_{(F_{0},G_{0})}z=F_{0}\mbox{Re}\int_{\Lambda}G_{0}^{*}Wdz+G_{0}\mbox{Re}\int_{\Lambda}F_{0}^{*}Wdz,
\]
where $\Lambda$ is a rectifiable curve going from $z_{0}$ upto $z_{1}$, in the complex plain. In particular, the $(F_{0},G_{0})$-integral of $\partial_{(F_{0},G_{0})}W$ reaches
\begin{equation}
\int_{z_{0}}^{z}\partial_{(F_{0},G_{0})}Wd_{F_{0},G_{0}}z=W-\phi (z_{0})F_{0}-\psi (z_{0})G_{0}.
\label{pre10}
\end{equation}
But according to the Theorem \ref{preFG}, the $(F_{0},G_{0})$-derivatives of $F_{0}$ and $G_{0}$ vanish identically, hence the equation (\ref{pre10}) can be considered the $(F_{0},G_{0})$-antiderivative of $\partial_{(F_{0},G_{0})}W$.
\end{definition}
\subsection{Formal Powers}
\begin{definition}
\label{preDEFFP}
The formal power $Z_{m}^{(n)}\left(a_{n},z_{0};z\right)$ belonging to the generating pair $(F_{m},G_{m})$, with formal exponent $n$, complex coefficient $a_{n}$, center at $z_{0}$, and depending upon the complex variable $z=x+iy$, is defined by the expression
\[
Z_{m}^{(n)}\left(a_{n},z_{0};z\right)=\lambda F_{m}+\mu G_{m},
\]
where $\lambda$ and $\mu$ are complex constants that fulfil the equality
\[
\lambda F_{m}(z_{0})+\mu G_{m}(z_{0})=a_{0}.
\]
The formal powers with higher formal exponents, are defined according to the recursive formulas
\begin{equation}
Z_{m}^{(n)}\left(a_{n},z_{0};z\right)=n\int_{z_{0}}^{z} Z_{m-1}^{(n-1)}\left(a_{n},z_{0};z\right)d_{(F_{m},G_{m})}z.
\label{pre11}
\end{equation}
Notice the integral operators at the right hand side of the last equality, are all $(F_{m},G_{m})$-antiderivatives. 
\end{definition}
\begin{remark}
\label{preFP}
The formal powers posses the following properties:
\begin{enumerate}
\item $Z_{m}^{(n)}\left(a_{n},z_{0};z\right)\rightarrow a_{n}\left(z-z_{0}\right)^{n}$ when $z\rightarrow z_{0}$.
\item All $Z_{m}^{(n)}\left(a_{n},z_{0};z\right)$ are $(F_{m},G_{m})$-pseudoanalytic.
\item If $a_{n}=a'_{n}+ia''_{n}$, where $a'_{n}$ and $a''_{n}$ are real constants, we will have that:
\begin{equation}
Z_{m}^{(n)}\left(a_{n},z_{0};z\right)=a'_{n}Z_{m}^{(n)}\left(1,z_{0};z\right)+a''_{n}Z_{m}^{(n)}\left(i,z_{0};z\right).
\label{pre12}
\end{equation}
\end{enumerate}
\end{remark}
\begin{theorem}
Every complex-valued function $W$, solution of the Vekua equation (\ref{pre05}), can be expanded in therms of the commonly called Taylor series in formal powers:
\begin{equation}
W=\sum_{n=0}^{\infty}Z^{(n)}\left(a_{n},z_{0};z\right),
\label{pre13} 
\end{equation}
where the absence of the subindex "$m$" indicates that all formal powers belong to the same generating pair.
\end{theorem}
\begin{remark}
\label{preremark}
Since every $W$, solution of (\ref{pre05}), accepts the representation (\ref{pre13}), it is possible to assert that (\ref{pre13}) is an analytic representation of the general solution for the Vekua equation (\ref{pre05}).
\end{remark}

\section{The Electrical Impedance Equation}
\label{two-dimensional}
As it has been previously posed in several works (see e.g. \cite{ckr}, \cite{kpa}, \cite{kra2005} and \cite{oct}), when the conductivity function $\sigma$ in (\ref{int00}) can be expressed by means of a separable-variables function
\begin{equation}
\sigma=\sigma_{1}(x)\sigma_{2}(y),
\label{eie00}
\end{equation}
when introducing the notations
\begin{eqnarray}
W=\sqrt{\sigma}\partial_{x}u-i\sqrt{\sigma}\partial_{y}u,\nonumber \\
p=\frac{\sqrt{\sigma_{2}}}{\sqrt{\sigma_{1}}};
\label{eie01}
\end{eqnarray}
the two-dimensional Electrical Impedance Equation (\ref{int00}) can be rewritten precisely as a Vekua equation of the form (\ref{pre08}). Moreover, its corresponding generating pair
\[
F_{0}=p,\ \ \ G_{0}=\frac{i}{p},
\]
is embedded into a periodic generating sequence, with period $2$, such that
\begin{enumerate}
\item \[F_{m}=\frac{\sqrt{\sigma_{2}}}{\sqrt{\sigma_{1}}},\ \ \ G_{m}=i\frac{\sqrt{\sigma_{1}}}{\sqrt{\sigma_{2}}};\] when $m$ is an even number, and
\item \[F_{m}=\sqrt{\sigma_{1}}\sqrt{\sigma_{2}},\ \ \ G_{m}=\frac{i}{\sqrt{\sigma_{1}}\sqrt{\sigma_{2}}};\] when $m$ is odd.
\end{enumerate}

Therefore, on the light of the statements posed in Definition \ref{preDEFFP}, the explicit generating sequence will allow us to approach the set of formal powers: 
\begin{equation}
\left\lbrace Z_{0}^{(n)}(1,0;z),Z_{0}^{(n)}(i,0;z)\right\rbrace_{n=0}^{N},
\label{eie02}
\end{equation}
within a bounded domain $\Omega$, and by virtue of the Remark \ref{preFP}, we will be able to approach any formal power $Z_{0}^{(n)}(a_n,0;z);\ n=0,\ 1,...,\ N$; at some certain point $z\in\Omega$.

Since the present work intends to be part of a novel theory for the Electrical Impedance Tomography problem, we will focus our attention into a classic domain $\Omega$: The unitary disk with center at $z_{0}=0$. We shall point out that all the results we will present, can be generalized for a wide class of bounded domains. However, it will be more comfortable by now to perform our analysis into this classical domain.

Hereafter, we will analyse the Dirichlet boundary value problem for the two-dimensional Electrical Impedance Equation (\ref{int00}).

\subsection{A complete orthonormal system}

In \cite{cck} V. Kravchenko \emph{et al.} posed a very important property of the formal powers, that will be the central column for our further discussions.
\begin{theorem}
\cite{cck} The set of real parts of the formal powers, with coefficients $1$ and $i$, corresponding to (\ref{pre08}), valued at the boundary $\Gamma$ of some domain $\Omega$:
\begin{equation}
\left\lbrace \mbox{Re}Z^{(n)}\left(1,0;z\right)\vert_{\Gamma},\mbox{Re}Z^{(n)}\left(i,0;z\right)\vert_{\Gamma}\right\rbrace_{n=0}^{\infty},
\label{eie03}
\end{equation}
constitute a complete system for the solutions of the Dirichlet boundary value problem of (\ref{int00}).
\end{theorem}

This is, any boundary condition $u\vert_{\Gamma}$ can be approached asymptotically by virtue of the linear combination of the elements belonging to (\ref{eie03}):
\[
\lim_{N\rightarrow\infty}\left(u\vert_{\Gamma}-\sum_{n=0}^{N}c^{(1)}_{n}\mbox{Re}Z^{(n)}\left(1,0;z\right)\vert_{\Gamma}-\sum_{n=0}^{N}c^{(i)}_{n}\mbox{Re}Z^{(n)}\left(i,0;z\right)\vert_{\Gamma}\right)=0,
\]
where the coefficients $\left\lbrace c_{n}^{(1)},c_{n}^{(i)}\right\rbrace_{n=0}^{\infty}$ are all real constants.

As a matter of fact, it is also possible to introduce an inner product for the set (\ref{int00}), as it was stated in \cite{cck}. In our particular case, this inner product will coincide with the Lebesgue integral, due to the boundary $\Gamma$ is the perimeter of the unitary circle. Hence, if $f(l)$ and $g(l)$ are two continuous real-valued functions defined over $\Gamma$ (this is $l\in (0,2\pi]$), their inner product will have the form
\begin{equation}
\left\langle f(l),g(l)\right\rangle=\int_{0}^{2\pi}f(l)g(l)dl.
\label{eie04}
\end{equation}

Therefore, we can always obtain an orthonormal set of functions $\lbrace u_{k}\rbrace_{k=0}^{\infty}$ using the elements of (\ref{eie03}), a property that will result most useful for our purposes. 

Specially talking about physical applications, by employing adequate numerical methods, as those described in \cite{bucio}; we will be able to approach the first $2N$ formal powers of the set (\ref{eie03}), from which it will arise an orthonormal the set of functions $\lbrace u_{k}\rbrace_{k=0}^{2N-1}$, ready to approach solutions for the Dirichlet problem of (\ref{int00}).

The effectiveness of these numerical calculations has been tried in several works (see \cite{cck}, \cite{ckr} and \cite{bsbp}). Still, it will be very convenient for us to analyse here a pair of particular examples, before paying all our attention into more general cases.

\section{Brief study of the Dirichlet boundary value problem, when $\sigma$ is separable-variables.}
\label{bsbvp}
\subsection{The case when $\sigma$ has exponential form.}
\label{exponential}
Let us consider a conductivity function of the form
\begin{equation}
\sigma=e^{x+y}.
\label{bs00}
\end{equation}
In order to obtain at least one analytic solution for the Electrical Impedance Equation, we will assume that (\ref{int00}) can be separated into a decoupled pair of Sturm-Liouville equations:  
\begin{eqnarray}
\partial_{x}\left(e^{x+y}\partial_{x} u\right)=0, \nonumber \\
\partial_{y}\left(e^{x+y}\partial_{y} u\right)=0.
\label{bs01}
\end{eqnarray}
It is remarkable that the modern Pseudoanalytic Function Theory has also provided new results for this classical kind of equations. Indeed, we could even approach the general solution for every equation of (\ref{bs01}), in terms of Taylor series in formal powers, employing the results posed in \cite{sturm}. Yet, at the moment, we are only searching for particular solutions.

For this purpose, let us introduce in the first equation of (\ref{bs01}) the auxiliary notation $u_{x}=\partial_{x}u$. We will have that
\[
\partial_{x}u_{x}+u_{x}=0,
\]
for which a particular solution is
\[
u_{x}=f(y)e^{-x},
\]
being $f(y)$ an arbitrary function depending upon $y$. Moreover, applying the antiderivative with respect to $x$ to the last expression, we can easily verify that
\begin{equation}
u=f(y)e^{-x}
\label{bs02}
\end{equation}
is a particular solution for the first equation of (\ref{bs01}). An identical procedure will show us that
\begin{equation}
u=g(x)e^{-y},
\label{bs03}
\end{equation}
where $g(x)$ is an arbitrary function of $x$, is a particular solution for the second. Once we combine (\ref{bs02}) and (\ref{bs03}), we will immediately obtain that
\begin{equation}
u=e^{-x-y}
\label{bs04}
\end{equation}
is an exact solution of the equation (\ref{int00}).

We have shown the full procedure because it will be useful for obtaining the required analytic solutions in every case we will study hereafter. 

Let us now summarise this result in a short Proposition.
\begin{proposition}
\label{prop01}
Let $\sigma=e^{x+y}$. Then the function $u=e^{-x-y}$, posed in (\ref{bs04}), will be a particular solution of (\ref{int00}).
\end{proposition}

Generating a boundary condition $u\vert_{\Gamma}$ from the particular solution shown in the previous Proposition, is a very simple task. Remembering that $\Gamma$ coincides with the perimeter of the unitary disk, we only have to consider $x=\cos\theta$ and $y=\sin\theta$, for $\theta\in [0,2\pi)$.

We can now employ the numerical methods detailed in \cite{bucio} for obtaining a system of $22$ formal powers, defined at the boundary $\Gamma$:
\begin{equation}
\left\lbrace Z_{0}^{(n)}(1,0;z)\vert_{\Gamma},Z_{0}^{(n)}(i,0;z)\vert_{\Gamma}\right\rbrace_{n=0}^{10},
\label{bs05}
\end{equation}
that correspond to the exponential conductivity $\sigma=e^{x+y}$. We construct each formal power to be composed by $1000$ complex values, associated with $1000$ equidistantly distributed points on $\Gamma$. Then, after applying a standard Gram-Schmitt orthonormalising process over the real parts of (\ref{bs05}), we will obtain a system of $21$ base functions $\left\lbrace u_{k}\right\rbrace_{k=0}^{20}$ (the apparent loss of one function can be explained using the expression (\ref{pre06}) introduced in Theorem \ref{theoremFG}, and the Definition \ref{preDEFFP}: The formal power $Z^{(0)}(i,0;z)$ is purely imaginary, thus its real part $\mbox{Re}Z^{(0)}(i,0;z)=0$).

Hence, every function $u_{k}$ will posses $1000$ real values along $\Gamma$. If needed, we can always use the well known cubic spline interpolling method on every $u_{k}$, to posses a set of continuous functions instead of a discrete one.

Our next step is to test the effectiveness of the set $\left\lbrace u_{k}\right\rbrace_{k=0}^{20}$ for approach solutions of the Dirichlet problem for (\ref{int00}), imposing the analytic solution (\ref{bs04}) as the boundary condition $u\vert_{\Gamma}$. Specifically, we will evaluate (\ref{bs04}) at the points 
\[
x=\cos\theta,\ \ y=\sin\theta;\ \ \theta=0,\left(\frac{2\pi}{21}\right),2\left(\frac{2\pi}{21}\right),3\left(\frac{2\pi}{21}\right),...,20\left(\frac{2\pi}{21}\right);
\] 
and the same will be done for every function of the set $\left\lbrace u_{k}\right\rbrace_{k=0}^{20}$. This will provide a basic linear system from which we will obtain $21$ constant coefficients $\left\lbrace \alpha_{k}\right\rbrace_{k=0}^{20}$ such that
\[
\sum_{k=0}^{20}\alpha_{k}u_{k}\sim u\vert_{\Gamma}.
\]

We shall point out that the number of evaluation points, located on the boundary $\Gamma$, does not necessarily have to coincide with the number $k$ of base elements. We selected to do this in behalf of simplicity, but other techniques are available if required. The Collocation Method, successfully employed in \cite{ckr}, is an example of the last assessment.

\begin{remark}
\label{remarkSequence} 
The $21$ elements of the set $\left\lbrace u_{k}\right\rbrace_{k=0}^{20}$ were obtained by applying the Gram-Schmitt orthonormalising process, considering first all the elements of the set 
\begin{equation}
\left\lbrace \mbox{Re}Z_{0}^{(n)}(1,0;z)\vert_{\Gamma}\right\rbrace_{n=0}^{10},
\label{bs06}
\end{equation}
and consequently the elements of
\begin{equation}
\left\lbrace \mbox{Re}Z_{0}^{(n)}(i,0;z)\vert_{\Gamma}\right\rbrace_{n=1}^{10}.
\label{bs07}
\end{equation}
\end{remark}
As expected, the absolute error $\mathcal{E}$ will be defined according to the Lebesgue norm:
\begin{equation}
\mathcal{E}=\left(\int_{0}^{2\pi}\left(\sum_{k=0}^{20}\alpha_{k}u_{k}-u\vert_{\Gamma}\right)^{2}dl\right)^{\frac{1}{2}}.
\label{bs08}
\end{equation}

\begin{figure}
\centering
\subfigure[Boundary condition and approached solution.]{
\includegraphics[scale=0.19]{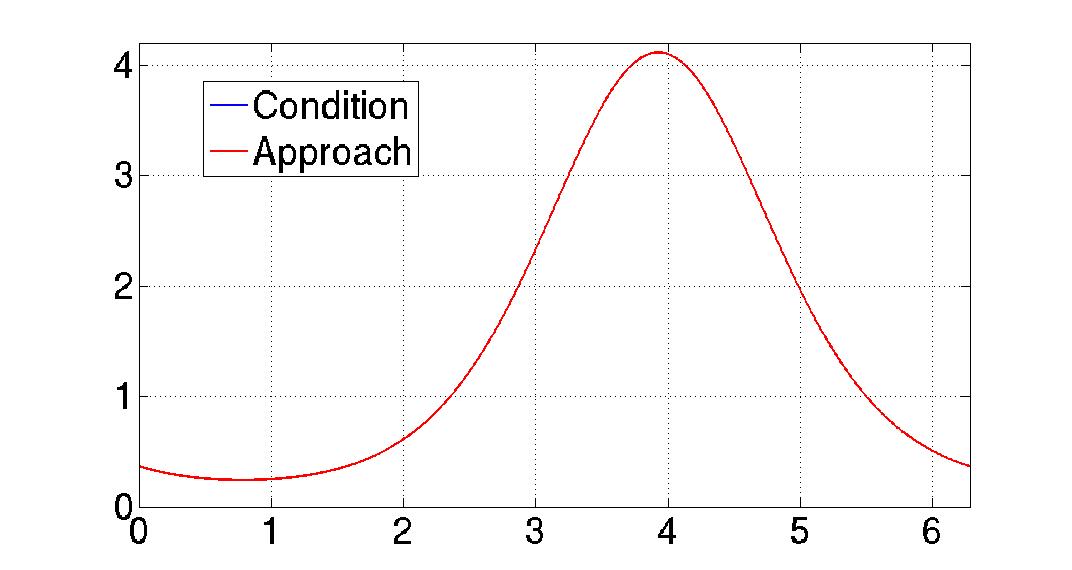}
\label{fig:bs00a}
}
\subfigure[Absolute values of the coefficients $\alpha_{k}$.]{
\includegraphics[scale=0.19]{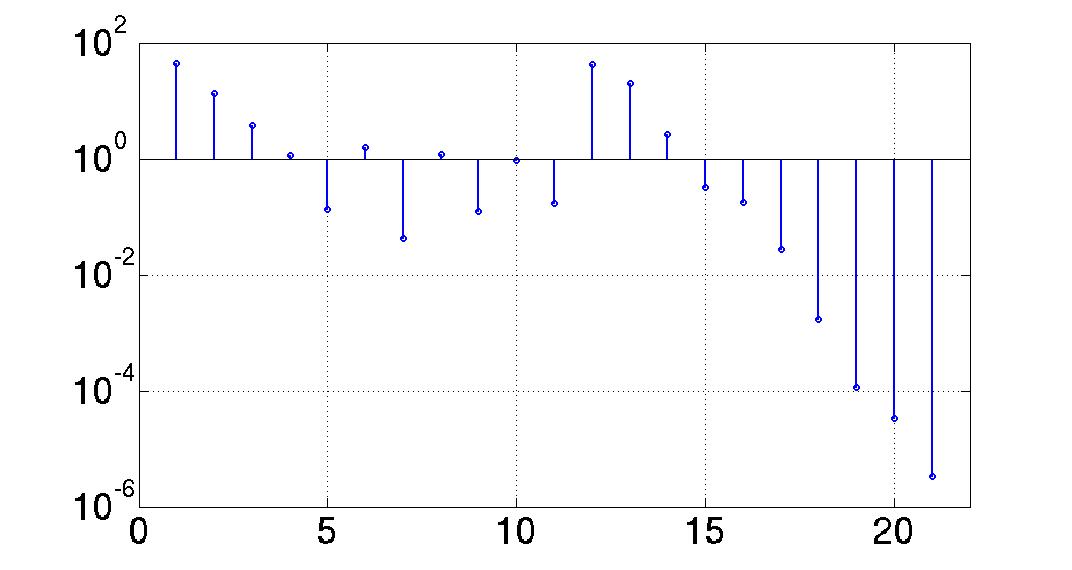}
\label{fig:bs00b}
}
\caption{Boundary condition and approached solution for the boundary value problem: Example 1.}
\end{figure}

The Figure \ref{fig:bs00a} shows the plot of the boundary condition $u\vert_{\Gamma}$ and the plot of the approached solution, being the horizontal axis the perimeter of the unitary disk. As the reader can appreciate, it is technically impossible to detect any difference in plain sight between these two graphics. That is why, when this case occurs, we will not display the illustration, focusing our attention into some of the most relevant coefficients $\alpha_{k}$, and the absolute error $\mathcal{E}$ defined in (\ref{bs08}).

\begin{table}
\centering
\caption{\label{tab:bs00}Values of the coefficients $\alpha_{k}$ corresponding to the boundary value problem with $\sigma$ of the form (\ref{bs00}).}
\begin{tabular}{| c | c | c | c | c | c | c | c |}
\hline
$\alpha_{0}$ & $\alpha_{1}$ & $\alpha_{2}$ & $\alpha_{3}$ & $\alpha_{11}$ & $\alpha_{12}$ & $\alpha_{13}$ & $\alpha_{14}$\\ \hline

43.708 & -13.435 & -3.713 & -1.152 & 41.760 & -19.752 & 2.693 & 0.327 \\ \hline

\end{tabular}
\end{table} 

The semilogarithmic Figure \ref{fig:bs00b} illustrates the absolute values of the $21$ coefficients $\alpha_{k}$ employed for approaching the condition $u\vert_{\Gamma}$, as well the Table \ref{tab:bs00} displays eight representative values of the set $\left\lbrace u_{k}\right\rbrace_{k=0}^{20}$, keeping in mind the Remark \ref{remarkSequence}, and rounded them to three decimal places. In other words, the first four values are the coefficients $\alpha_{k}$ corresponding to the orthonormal functions upcoming from the set (\ref{bs05}), whereas the remaining values correspond to the coefficients of the orthonormal functions obtained from (\ref{bs06}).

From Figure \ref{fig:bs00b}, we could venture to say that the elements shown in Table \ref{tab:bs00} are among the most significant for the approach.

The absolute error introduced in (\ref{bs08}), and also rounded to three decimal places, is $\mathcal{E}=2.006\times 10^{-8}$. Its integral expression was approached by the trapezoidal method, considering $1000$ equidistant segments among the interval $[0,2\pi)$. This parameters will be kept for further calculations.

\subsection{The case when $\sigma$ has a Lorentzian form}
\label{lorentzianSV}

This example was selected because its exhibited numerical properties, that will be properly shown soon after. Beside, this conductivity could be situated very close of real Medical Imaging examples. Following identical logic steps to those exposed in the last section, we can pose the following statement. 
\begin{proposition}
\label{prop02}
Let
\begin{equation}
\sigma=\left(\frac{1}{x^2+0.1}\right)\left(\frac{1}{y^2+0.1}\right).
\label{bs09}
\end{equation} 
Then the function
\begin{equation}
u=\frac{x^3+y^3}{3}+0.1\left(x+y\right),
\label{bs10}
\end{equation}
will be a particular solution of (\ref{int00}).
\end{proposition}

As in the previous case, we will use (\ref{bs10}) to construct the boundary condition $u\vert_{\Gamma}$, we will approach a set of orthonormal functions $\left\lbrace u_{k}\right\rbrace_{k=0}^{20}$, and we will calculate the coefficients $\alpha_{k}$ to approach the boundary condition.

\begin{figure}
\centering
\includegraphics[scale=0.19]{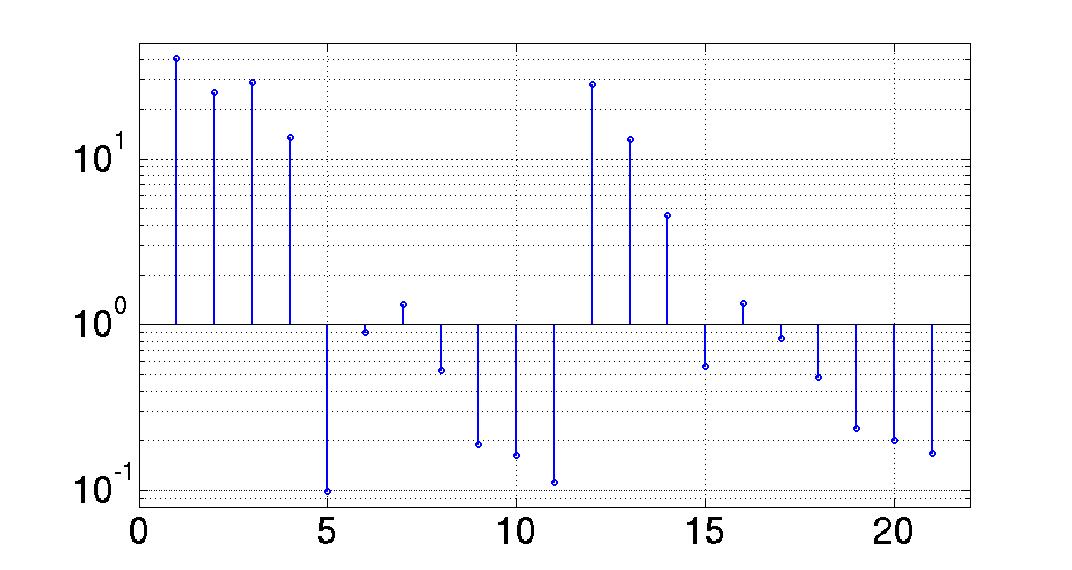}
\label{figbs01}
\caption{Absolute values of the coefficients $\alpha_{k}$, corresponding to the case of the Lorentzian conductivity (\ref{bs09}).}
\end{figure}

It is evident from the Figure 2 that the values of the significant coefficients $\alpha_{k}$ are higher than those obtained for the exponential case, but the number of significant coefficients remains, somehow, the same. This can be corroborated watching the contents of Table \ref{tab:bs01}.

\begin{table}
\centering
\caption{\label{tab:bs01}Values of the coefficients $\alpha_{k}$ corresponding to the boundary value problem with $\sigma$ of the from (\ref{bs09}).}
\begin{tabular}{| c | c | c | c | c | c | c | c |}
\hline
$\alpha_{0}$ & $\alpha_{1}$ & $\alpha_{2}$ & $\alpha_{3}$ & $\alpha_{11}$ & $\alpha_{12}$ & $\alpha_{13}$ & $\alpha_{14}$\\ \hline
 40.216 & -25.102 & -28.884 & 13.505 &  28.106 & -13.119 &  4.576 &  -0.565 \\ \hline
\end{tabular}
\end{table}  

For this case, the absolute error was $\mathcal{E}= 2.15\times 10^{-2}$. The error is considerably bigger than the one obtained in the previous subsection. Nevertheless, there is no need to illustrate the plots of the condition and of the reconstruction, since not any difference can be appreciated. In the opinion of the authors, the approached solution can still be considered adequate.

\section{Analysis of boundary value problems when the conductivity is not originally a separable-variables function.}

Basically, all results showed above are valid if and only if the conductivity function $\sigma$ can be expressed in terms of a separable variables function of the form (\ref{eie00}). This, of course, will seldom happen in physical applications, but it seems to be an alternative to approach any given conductivity function in terms of a piecewise separable-variables function, that might be useful for practical applications.

It is not clear if the proof of completeness for the set of formal powers within bounded domains, elegantly posed by Kravchenko \emph{et al.} in \cite{cck}, can be extended for the case of a piecewise separable-variables function, nor it is the intention of this work to try to do so. Yet, the numerical results indicate that this alternative is worth of consideration, and the following paragraphs will intend to support this assertion.

The idea for introducing a piecewise separable-variables function was first presented in \cite{oct}, and a basic characterization was posed in \cite{bath2011} and \cite{wce2011}. Here, the proposal will be analysed by studying six different examples. It will be used to approach solutions for the two cases previously studied, and it will be employed to analyse four cases where the conductivities are not, by definition, separable-variables functions. Just as performed before, exact solutions will be used for generating the boundary conditions to achieve in each example.

\subsection{Construction of a piecewise separable-variables conductivity function}
\label{piecewise}

Consider a bounded domain $\Omega$ (in this case the unitary disk), and divide it into a finite number of subsections, taking care that the point to be considered the center of the formal powers $z_{0}$ (see Definition \ref{preDEFFP}) does not reside onto the boundary of two or more subsections. In behalf of simplicity, let us make the division by employing a finite set of parallel lines to the $y$-axis, equidistant one to each other, and let us locate $z_{0}=0$.

Supposing that the values of the electrical conductivity are defined at every point inside the domain $\Omega$, let us trace a straight line within every subsection, watching out that does not intersect the bounding parallel lines of its corresponding subsection, inside $\Omega$. Indeed, such lines can simply be parallel to the bounding ones.

The next step is to collect a finite set of values over every line that crosses the subsections. For every crossing line, the quantity of collected values must be large enough to warrant that an interpolling process (in our case cubic splines) will adequately approach all the remaining conductivity values over the line. 

Since we already assumed that every crossing line will be parallel to the subsection-bounding lines, and in consequence to the $y$-axis, all collected points corresponding to the same crossing line will posses the same $x$-coordinate. Let us now propose that the conductivity inside every subsection can be represented according to the expression
\begin{equation}
\sigma=\frac{x+K}{\chi+K}\cdot f(y),
\label{abv00}
\end{equation}
where $\chi$ denotes the $x$-coordinate that is common to all points along the crossing line, $f(y)$ is an interpolling function that approaches the values of the conductivity collected over the line, and K is a positive real constant such that $x+K\neq 0$ within the subsection. 

From this point of view, and supposing we have $M$ subsections, the conductivity $\sigma$ inside the bounded domain $\Omega$ can be approached by means of the piecewise function

\begin{displaymath}
   \sigma (x,y) = \left\{
     \begin{array}{lr}
       \frac{x+K_{1}}{\chi_{1}+K_{1}}\cdot f_{1}(y) & : x \in [x_{(1)},x_{(2)});\\
       \frac{x+K_{2}}{\chi_{2}+K_{2}}\cdot f_{2}(y) & : x \in [x_{(2)},x_{(3)});\\
       \cdots & \\
       \frac{x+K_{M}}{\chi_{M}+K_{M}}\cdot f_{M}(y) & : x \in [x_{(M)},x_{(M+1)}].
     \end{array}
   \right.   
\end{displaymath}
\begin{equation}
\label{abv01}
\end{equation}
Here $x_{(1)}$ represents the first $x$-coordinate found within the domain $\Omega$ when broaching the $x$-axis from $-\infty$ upto $\infty$, whereas $x_{(M)}$ represents the last one. The pairs of coordinates $(x_{(j)},x_{(j+1)})$, where $\ \ j=0,1,...,M+1$; represent the common $x$-coordinates belonging to every pair of lines delimiting the subsections. It is evident that the piecewise function (\ref{abv01}) is separable-variables.

According to the Section \ref{two-dimensional}, it immediately follows that

\begin{displaymath}
   F_{0} = \left\{
     \begin{array}{lr}
       \left(\frac{\chi_{1}+K_{1}}{x+K_{1}}\cdot f_{1}(y)\right)^{\frac{1}{2}} & : x \in [x_{(1)},x_{(2)});\\
       \left(\frac{\chi_{2}+K_{2}}{x+K_{2}}\cdot f_{2}(y)\right)^{\frac{1}{2}} & : x \in [x_{(2)},x_{(3)});\\
       \cdots & \\
       \left(\frac{\chi_{M}+K_{M}}{x+K_{M}}\cdot f_{M}(y)\right)^{\frac{1}{2}} & : x \in [x_{(M)},x_{(M+1)}].
     \end{array}
   \right.   
\end{displaymath}
Whereas
\begin{displaymath}
   G_{0} = \left\{
     \begin{array}{lr}
       i\left(\frac{x+K_{1}}{\chi_{1}+K_{1}}\cdot \frac{1}{f_{1}(y)} \right)^{\frac{1}{2}} & : x \in [x_{(1)},x_{(2)});\\
       i\left(\frac{x+K_{2}}{\chi_{2}+K_{2}}\cdot \frac{1}{f_{2}(y)}\right)^{\frac{1}{2}} & : x \in [x_{(2)},x_{(3)});\\
       \cdots & \\
       i\left(\frac{x+K_{M}}{\chi_{M}+K_{M}}\cdot \frac{1}{f_{M}(y)}\right)^{\frac{1}{2}} & : x \in [x_{(M)},x_{(M+1)}].
     \end{array}
   \right.   
\end{displaymath}

For the generating pair $(F_{1},G_{1})$ we will simply have 
\[
F_{1}=\sqrt{\sigma},\ \ \ G_{1}=\frac{i}{\sqrt{\sigma}};
\] 
being $\sigma$ the piecewise separable-variables function introduced in (\ref{abv01}). These are the generating pairs $(F_{0},G_{0})$ and $(F_{1},G_{1})$ that we will employ for the numerical calculations.

\subsection{The separable-variables exponential case}

Let us consider again the exponential conductivity function introduced in (\ref{bs00}): $e^{x+y}$. We will use it for approaching the piecewise-separable variables generating pairs $(F_{0},G_{0})$ and $(F_{1},G_{1})$ declared in the Subsection above.

Hereafter, we will consider $M=1001$ subsections, produced by equidistantly parallel lines to the $y$-axis, and the parallel-crossing lines will be placed at the very middle of every subsection. Also, for every subsection, we will take $1000$ samples of conductivity, collected from an equal number of points equidistantly distributed over the crossing line of the subsection. Finally, we will assume $K_{1}=K_{2}=...=K_{M}=60$.

Following the methodology posed in Section \ref{bsbvp}, we can build an orthonormal system $\left\lbrace u_{k}\right\rbrace_{k=0}^{20}$, to approach the boundary condition $u\vert_{\Gamma}$, raised from the corresponding particular solution of (\ref{int00}) $u=e^{-x-y}$.

\begin{figure}
\centering
\includegraphics[scale=0.19]{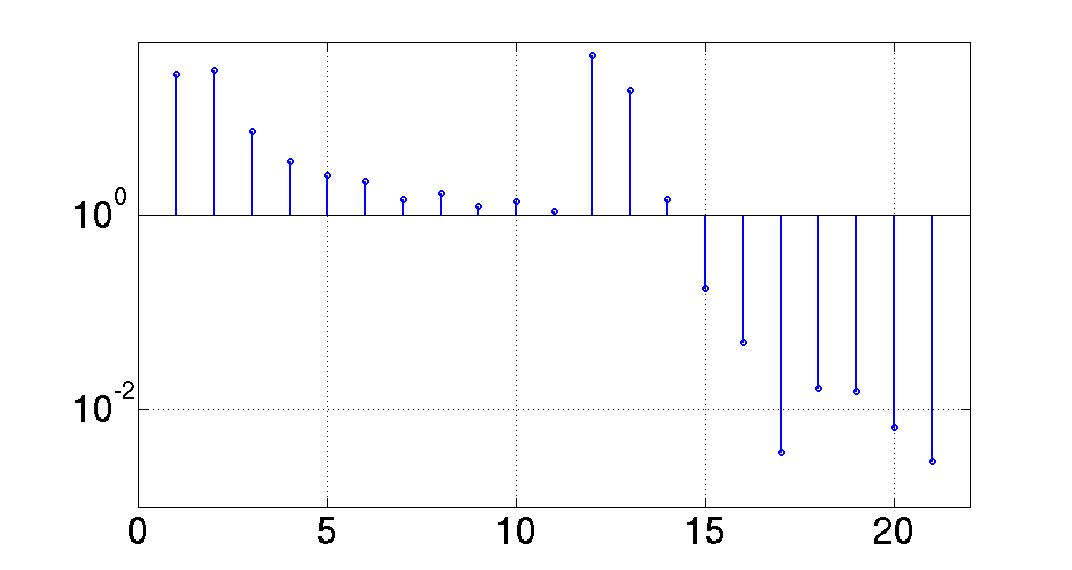}
\label{fig:abv00}
\caption{Absolute values of the coefficients $\alpha_{k}$, corresponding to the case of the piecewise conductivity resulting from (\ref{bs00}).}
\end{figure}

The Figure 3 shows that the number of significant coefficients $\alpha_k$ has been increased. It seems that more than ten are among the most relevant. Nevertheless, many of these values have very similar magnitudes, therefore in the Table \ref{tab:abv00} we will only show eight values, as is has be done before. The absolute error has also grown, being $\mathcal{E}=3.4\times 10^{-3}$. Still, not any difference is noticed when plotting together the boundary condition and the approached solution.

\begin{table}
\centering
\caption{\label{tab:abv00}Values of the coefficients $\alpha_{k}$ corresponding to the case of the piecewise conductivity resulting from (\ref{bs00}).}
\begin{tabular}{| c | c | c | c | c | c | c | c |}
\hline
$\alpha_{0}$ & $\alpha_{1}$ & $\alpha_{2}$ & $\alpha_{3}$ & $\alpha_{11}$ & $\alpha_{12}$ & $\alpha_{13}$ & $\alpha_{14}$\\ \hline

 28.539 &  -31.624 &  7.445 &  3.649 &  44.349 & -19.430 &   1.479 &  0.179 \\ \hline

\end{tabular}
\end{table}  

\subsection{The separable-variables Lorentzian case}

We now consider the conductivity function (\ref{bs09}) once more, performing the numerical calculations with all the details provided in the previous Subsection, and employing the piecewise separable-variables conductivity function obtained from (\ref{bs09}). The boundary condition rises from the exact solution (\ref{bs10}) of (\ref{int00}).

\begin{figure}
\centering
\subfigure[Boundary condition and approached solution.]{
\includegraphics[scale=0.19]{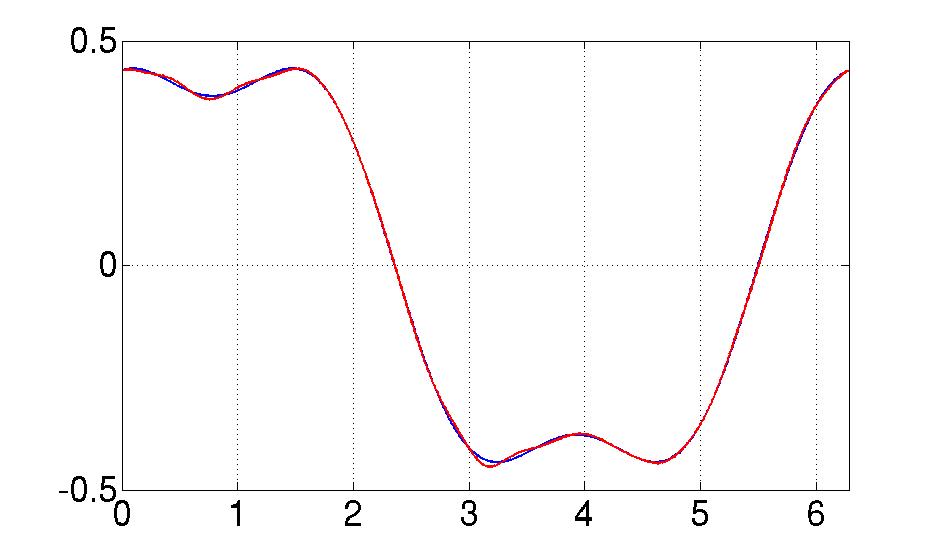}
\label{fig:abv01a}
}
\subfigure[Absolute values of the coefficients $\alpha_{k}$.]{
\includegraphics[scale=0.19]{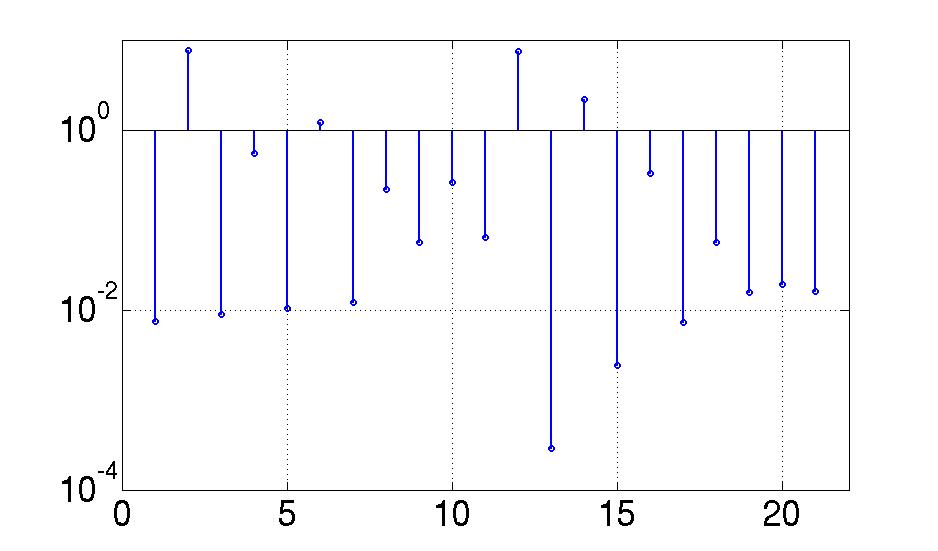}
\label{fig:abv01b}
}
\caption{Results of the calculations employing the piecewise-separable variable function for the Lorentzian conductivity}
\end{figure}

This is the first case where it is possible to detect a clear difference between the plot corresponding to the boundary condition (blue line) and the approached solution (red line), as it is illustrated in Figure \ref{fig:abv01a}. Moreover, even the number of significant coefficients remains low (about four only), they do not appear following the same sequence that tables shown before, as we can appreciate in Figure \ref{fig:abv01b}. Please examine carefully the numbers of the subindexes contained in Table 4. About the absolute error, we have $\mathcal{E}=10.2\times 10^{-3}$.

\begin{table}
\centering
\caption{\label{tab:abv01}Values of the coefficients $\alpha_{k}$ corresponding to the case of the piecewise conductivity resulting from (\ref{bs09}).}

\begin{tabular}{| c | c | c | c | c | c | c | c |}
\hline
$\alpha_{1}$ & $\alpha_{3}$ & $\alpha_{5}$ & $\alpha_{9}$ & $\alpha_{12}$ & $\alpha_{14}$ & $\alpha_{16}$ & $\alpha_{18}$\\ \hline

 7.9239 &   0.565 &  -1.225 &  -0.266 &  -7.724 &  2.220 &   -0.330 &  0.057 \\ \hline

\end{tabular}
\end{table}  

\subsection{The non-separable variables exponential case.}

The following four examples intend to show that the methodology for approaching piecewise separable-variables conductivity functions posed in Subsection \ref{piecewise}, can be trustful even when the conductivity function was not originally separable-variables.

We will aboard this task considering first an exponential conductivity.
\begin{proposition}
Let the conductivity function 
\begin{equation}
\sigma=e^{xy}.
\label{abv02}
\end{equation}
Then, a particular solution of (\ref{int00}) will be
\begin{equation}
u=e^{-xy}.
\label{abv03}
\end{equation}
\end{proposition}

Once more, we use (\ref{abv02}) for creating a piecewise separable-variables function, and we impose (\ref{abv03}) as the boundary condition. The obtained results are the following.

\begin{figure}
\centering
\includegraphics[scale=0.19]{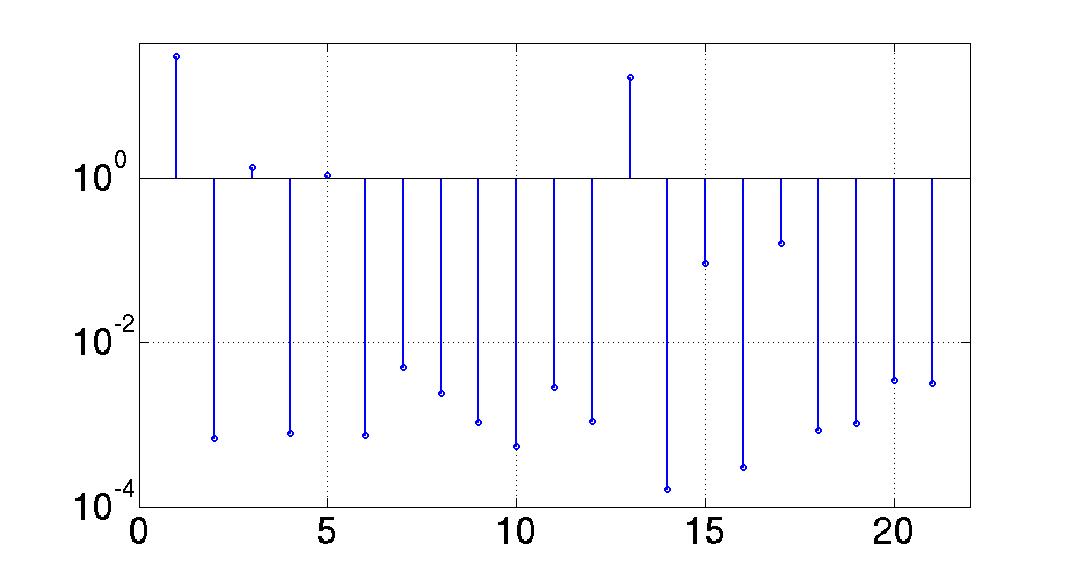}
\label{fig:abv03}
\caption{Absolute values of the coefficients $\alpha_{k}$, corresponding to the case of the piecewise conductivity resulting from (\ref{abv02}).}
\end{figure}

It is not possible to notice any difference between the graphic of the boundary condition and the plot of the numerical approach. That is why we only show the semilogarithmic graphic of the coefficients $\alpha_{k}$ in the Figure 5. For this case, we only identify two relevant coefficients, but we still show eight values in the Table \ref{tab:abv03}, as usual. The approached absolute error for this case is $\mathcal{E}=8.948\times 10^{-4}$.

\begin{table}
\centering
\caption{\label{tab:abv03}Values of the coefficients $\alpha_{k}$ corresponding to the case of the piecewise conductivity resulting from (\ref{abv02}).}

\begin{tabular}{| c | c | c | c | c | c | c | c |}
\hline
$\alpha_{0}$ & $\alpha_{1}$ & $\alpha_{2}$ & $\alpha_{3}$ & $\alpha_{11}$ & $\alpha_{12}$ & $\alpha_{13}$ & $\alpha_{14}$\\ \hline

 31.150 &   0.001 &   -1.378 &   -0.001 &  0.001 &  17.105 &    0.001 &   -0.093 \\ \hline

\end{tabular}
\end{table}  

\subsection{The non-separable variables Lorentzian case.}

\begin{proposition}
Let the conductivity function have the form
\begin{equation}
\sigma=\frac{1}{\left(x+y\right)^{2}+1}.
\label{abv04}
\end{equation}
An exact solution for the equation (\ref{int00}) is
\begin{equation}
u=\frac{\left(x+y\right)^{3}}{3}+x+y.
\label{abv05}
\end{equation}
\end{proposition}

We construct the piecewise separable-variables conductivity employing (\ref{abv04}), and we introduce the boundary condition using (\ref{abv05}). For this example we neither find any remarkable difference between the graphics of the condition and the approach. The Figure 6 displays the magnitude of the coefficients $\alpha_{k}$. We detect only four relevant coefficients, as assented in Table \ref{tab:abv05}. The absolute error is $\mathcal{E}=1.4\times 10^{-3}$.

\begin{figure}
\centering
\includegraphics[scale=0.19]{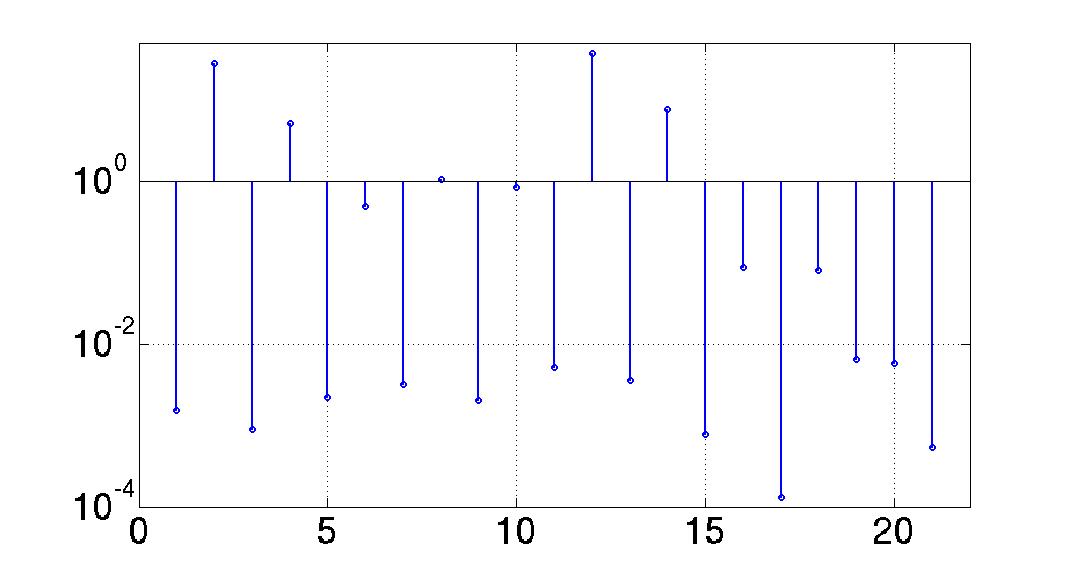}
\label{fig:abv06}
\caption{Absolute values of the coefficients $\alpha_{k}$, corresponding to the case of the piecewise conductivity resulting from (\ref{abv04}).}
\end{figure}

\begin{table}
\centering
\caption{\label{tab:abv05}Values of the coefficients $\alpha_{k}$ corresponding to the case of the piecewise conductivity resulting from (\ref{abv04}).}

\begin{tabular}{| c | c | c | c | c | c | c | c |}
\hline
$\alpha_{0}$ & $\alpha_{1}$ & $\alpha_{2}$ & $\alpha_{3}$ & $\alpha_{11}$ & $\alpha_{12}$ & $\alpha_{13}$ & $\alpha_{14}$\\ \hline

 0.002 &   28.071 &    -0.001 &    -5.087 &  -37.473 &  -0.004 &    -7.575 &   0.001 \\ \hline
 
\end{tabular}
\end{table}

\subsection{The non-separable variables polynomial case.}

\begin{proposition}
Let us assume the conductivity function to have the form
\begin{equation}
\sigma=x+y+10,
\label{abv06}
\end{equation}
thus the function
\begin{equation}
u=\mbox{ln}\left(x+y+10\right),
\label{abv07}
\end{equation}
will be a solution of (\ref{int00}).
\end{proposition}

As usual, first it comes the approaching of the piecewise separable-variables conductivity based upon (\ref{abv06}), and latter we stablish the boundary condition according to (\ref{abv07}). One more time, there is nothing to enhance from the comparative graphic of the condition and the approach. The absolute values of the coefficients are displayed in Figure 7. Nevertheless this figure suggest that there are at least four coefficients playing a relevant role in the approach, we can venture to point out that perhaps only three are really taking a fundamental part. The Table 7 will show why we estimate that. We shall ask again the reader to observe the numbers of the subindexes. The total error for this case is $\mathcal{E}=9.8\times 10^{-3}$.

\begin{figure}
\centering
\includegraphics[scale=0.19]{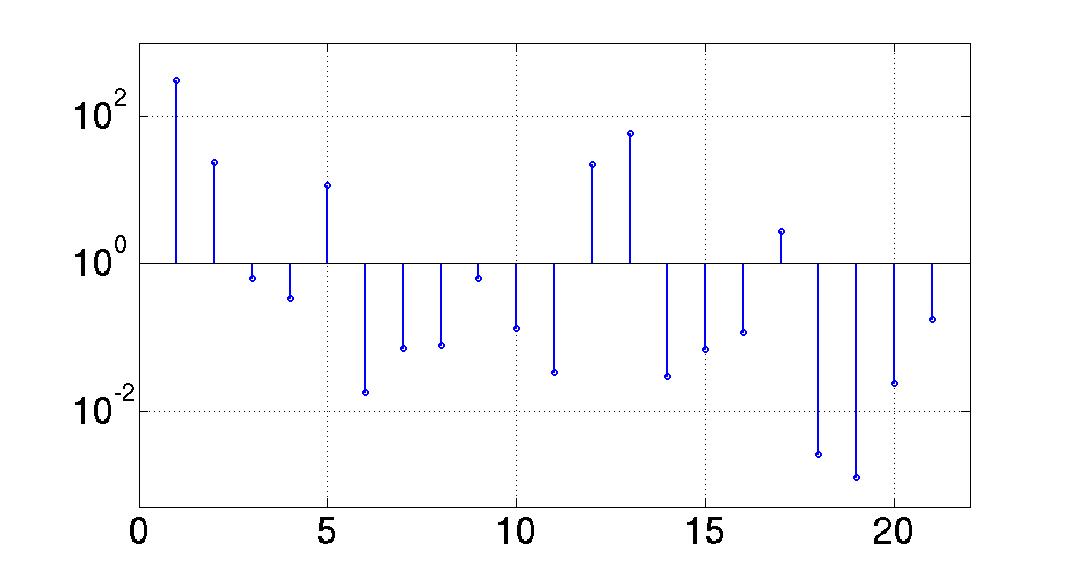}
\label{fig:abv07}
\caption{Absolute values of the coefficients $\alpha_{k}$, corresponding to the case of the piecewise conductivity resulting from (\ref{abv06}).}
\end{figure}

\begin{table}
\centering
\caption{\label{tab:abv07}Values of the coefficients $\alpha_{k}$ corresponding to the case of the piecewise conductivity resulting from (\ref{abv06}).}

\begin{tabular}{| c | c | c | c | c | c | c | c |}
\hline
$\alpha_{0}$ & $\alpha_{1}$ & $\alpha_{2}$ & $\alpha_{4}$ & $\alpha_{11}$ & $\alpha_{12}$ & $\alpha_{13}$ & $\alpha_{14}$\\ \hline

 310.378 &   24.207 &     -0.638 &    11.684 &  -22.382 &  -58.599 &    -0.030 &    0.068 \\ \hline

\end{tabular}
\end{table}

\subsection{The non-separable variables sinusoidal case.}

This case is the last one of our characterization, and we shall remark that it could offer interesting challenges for a more detailed examination, due to the possibility of numerical indetermination contained into its particular solution. At this moment, we will consider a basic example.

\begin{proposition}
Let us consider the sinusoidal conductivity
\begin{equation}
\sigma=1+\sin xy.
\label{abv08}
\end{equation}
We can verify by direct substitution that the function
\begin{equation}
u=\left(\tan\left(\frac{xy}{2}\right)+1\right)^{-1},
\label{abv09}
\end{equation}
is a solution of (\ref{int00}).
\end{proposition}

For the last time, and noticing the conductivity (\ref{abv08}) does not vanish inside the unitary disk, we will employ it for approaching the piecewise conductivity function, and (\ref{abv09}) for constructing the boundary condition. The Figure 8 illustrates the absolute values of $\alpha_{k}$, whereas the Table 8 shows us that, for this example, only two coefficients take an important part into the approach. The total error is $\mathcal{E}=7.694\times 10^{-4}$.

\begin{figure}
\centering
\includegraphics[scale=0.19]{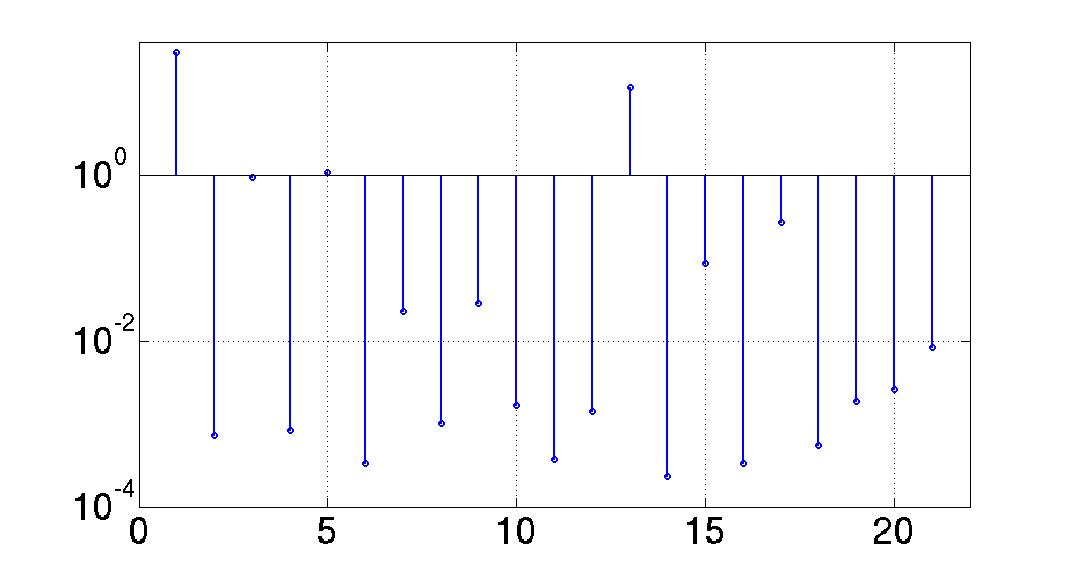}
\label{fig:abv08}
\caption{Absolute values of the coefficients $\alpha_{k}$, corresponding to the case of the piecewise conductivity resulting from (\ref{abv08}).}
\end{figure}

\begin{table}
\caption{\label{tab:abv08}Values of the coefficients $\alpha_{k}$ corresponding to the case of the piecewise conductivity resulting from (\ref{abv08}).}

\begin{tabular}{| c | c | c | c | c | c | c | c |}
\hline
$\alpha_{0}$ & $\alpha_{1}$ & $\alpha_{2}$ & $\alpha_{4}$ & $\alpha_{11}$ & $\alpha_{12}$ & $\alpha_{13}$ & $\alpha_{14}$\\ \hline

 31.1336 &    0.0007 &     -0.9532 &    -0.0008 &   0.0014 &  11.5636 &    0.0002 &    -0.0881 \\ \hline

\end{tabular}
\end{table}

\section{Conclusions}

Based upon the results showed in the previous sections, we consider that the basic characterization of the new numerical method
for approaching solutions of the Electrical Impedance Equation is acceptable, since it has provided the proper information for performing new experiments on a wider class of conductivity functions that, by definition, are not separable-variables, as well as for starting the analysis of geometrical conductivity distributions, whose exact mathematical representation are, in general, unknown.

This second class of conductivity distributions is among the most interesting for several branches of Experimental Physics, as it is the Medical Imaging. But it is clear that the characterization of the new method over this class of conductivities, can only be adequately achieved by comparing its numerical results with those provided by some of the finest variations of the Finite Element Method, that have proved to be the best tool to approach solutions for Dirichlet boundary value problems in the plain.

The selection of the proper examples for these tests could well be worth of a complete paper. Anyway, it is absolutely necessary to perform those comparatives.

We must also emphasize that we have studied only one class of piecewise separable-variables conductivity approach. Indeed, to try again the posed examples, and as much of their variations as it is possible, using other ideas for obtaining separable-variables expressions at every subsection, will enrich our understanding of this new technique from the mathematical point of view.

On the same direction, it is also necessary to formally extend the proof of completeness provided in \cite{cck}, if such was possible, for piecewise separable-variables functions. About this, the authors do not detect at the present any immediate way to start accomplishing the task.

And we still have to take into account all material related to the Computational Complexity of the new posed numerical methods, that was completely out of the scope of the current paper. Yet, this path could be somehow clearer to start exploring, if we keep in mind that our ultimate objective is to apply the methods for Medical Imaging problems, so it will be obligatory to include the computational cost and time as two parameters for evaluating the technical efficiency of the new methods.
This because the known algorithms for approaching solutions for the inverse problem, are based on the recursive resolution of the direct Dirichlet boundary value problem.

Still, we are certain to be searching in the correct direction, because the techniques provided by the Pseudoanalytic Function Theory, classical \cite{bers}\cite{vekua}, and modern \cite{kpa}, are proving to be specially powerful in Mathematical Physics, e.g., these techniques could well prove useful in an assortment relevant applications, as it is the Fourier transform of the two-dimensional Fokker-Planck equation, studied in \cite{hijar}, as well as some certain special cases of the diffusion equation employed into hydrodynamic fluctuations \cite{hijar2}.
 
\begin{acknowledgement}
The authors would like to acknowledge the support of CONACyT project 106722, Mexico. M. P. Ramirez T. thanks the support of HILMA S.A. de C.V., Mexico.
\end{acknowledgement}

\end{document}